\documentclass[%
 aip,
 jmp,%
 amsmath,amssymb,
 reprint,%
]{revtex4-2}

\usepackage{graphicx}
\usepackage{dcolumn}
\usepackage{amssymb,amsmath}
\usepackage{bbm}
\usepackage{booktabs}
\usepackage{siunitx}
\usepackage[hidelinks]{hyperref}
\usepackage{xcolor}

\usepackage{courier}
\usepackage{listings}

\renewcommand{\vec}[1]{\boldsymbol{#1}}
\newcommand{\vnabla}{\vec{\nabla}}

\newcommand{\dx}{\,\text{d}\vec{x}}
\newcommand{\ds}{\,\text{d}\vec{s}}

\begin{document}

\title{NeuralMag: an open-source nodal finite-diﬀerence code for inverse micromagnetics}

\author{C. Abert}%
\email{claas.abert@univie.ac.at}%
\affiliation{Faculty of Physics, University of Vienna, 1090 Vienna, Austria}%
\author{F. Bruckner}%
\affiliation{Faculty of Physics, University of Vienna, 1090 Vienna, Austria}%
\author{A. Voronov}%
\affiliation{Faculty of Physics, University of Vienna, 1090 Vienna, Austria}%
\author{M. Lang}%
\affiliation{Max Planck Institute for the Structure and Dynamics of Matter, 22761 Hamburg, Germany}%
\affiliation{Center for Free-Electron Laser Science, Luruper Chaussee 149, 22761 Hamburg, Germany}%
\author{S. A. Pathak}%
\affiliation{Max Planck Institute for the Structure and Dynamics of Matter, 22761 Hamburg, Germany}%
\affiliation{Center for Free-Electron Laser Science, Luruper Chaussee 149, 22761 Hamburg, Germany}%
\author{S. Holt}%
\affiliation{Max Planck Institute for the Structure and Dynamics of Matter, 22761 Hamburg, Germany}%
\affiliation{Center for Free-Electron Laser Science, Luruper Chaussee 149, 22761 Hamburg, Germany}%
\author{R. Kraft}%
\affiliation{Faculty of Physics, University of Vienna, 1090 Vienna, Austria}%
\author{R. Allayarov}%
\affiliation{Faculty of Physics, University of Vienna, 1090 Vienna, Austria}%
\author{P. Flauger}%
\affiliation{Faculty of Physics, University of Vienna, 1090 Vienna, Austria}%
\author{S. Koraltan}%
\affiliation{Institute of Applied Physics, TU Wien, 1040 Vienna, Austria}%
\author{T. Schrefl}%
\affiliation{Department for Integrated Sensor Systems, University for Continuing Education Krems, 3500 Krems, Austria}%
\author{A. Chumak}%
\affiliation{Faculty of Physics, University of Vienna, 1090 Vienna, Austria}%
\author{H. Fangohr}%
\affiliation{Max Planck Institute for the Structure and Dynamics of Matter, 22761 Hamburg, Germany}%
\affiliation{Center for Free-Electron Laser Science, Luruper Chaussee 149, 22761 Hamburg, Germany}%
\affiliation{University of Southampton, SO17 1BJ Southampton, United Kingdom}%
\author{D. Suess}%
\affiliation{Faculty of Physics, University of Vienna, 1090 Vienna, Austria}%

\date{\today}%

\begin{abstract}
We present NeuralMag, a flexible and high-performance open-source Python library for micromagnetic simulations.
NeuralMag leverages modern machine learning frameworks, such as PyTorch and JAX, to perform efficient tensor operations on various parallel hardware, including CPUs, GPUs, and TPUs.
The library implements a novel nodal finite-difference discretization scheme that provides improved accuracy over traditional finite-difference methods without increasing computational complexity.
NeuralMag is particularly well-suited for solving inverse problems, especially those with time-dependent objectives, thanks to its automatic differentiation capabilities.
Performance benchmarks show that NeuralMag is competitive with state-of-the-art simulation codes while offering enhanced flexibility through its Python interface and integration with high-level computational backends.
\end{abstract}

\maketitle

\section{Introduction}
Micromagnetic simulations are a fundamental tool in the study of magnetization dynamics and play a crucial role in understanding and designing magnetic materials and devices.
These simulations model the behavior of magnetic and magnonic systems at the nanoscale, providing insight into phenomena such as domain wall motion, magnetization reversal, and spin wave propagation.
The field relies on various computational methods, with finite-difference and finite-element schemes being widely used.
Notable examples of established finite-difference codes include OOMMF\cite{donahue1999oommf} and fidimag\cite{bisotti2020fidimag} for CPU-based simulations and mumax3\cite{vansteenkiste2014design} and magnum.np\cite{bruckner2023magnum} for GPU-accelerated simulations.
Finite-element-based methods, such as those implemented in NMag\cite{fischbacher2007systematic}, Tetramag\cite{kakay2010speedup}, FastMag\cite{chang2011fastmag}, FinMag\cite{bisotti2018finmag}, and magnum.fe\cite{abert2013magnum}, provide greater flexibility in handling complex geometries but can be computationally more expensive.

In addition to standard micromagnetic simulations, inverse problems have attracted considerable attention in recent years.
These problems involve determining the optimal parameters --- such as material properties, external fields, or device geometries --- that lead to a desired magnetic configuration or device functionality.
A significant body of work has focused on inverse modeling of the demagnetization field, a static inverse problem.
This has been particularly useful in the context of magnetic 3D printing, where topology optimization techniques are employed to design optimal material layouts, and the inverse modeling is used to infer the magnetization configuration of printed samples\cite{bruckner2017solving,abert2017fast,huber2017topology}.

More recently, research in the emerging field of inverse magnonics has gained momentum, focusing on optimizing the functionality of magnonic devices.
Magnonics uses spin waves (magnons) for information processing, and designing efficient magnonic devices poses complex nonlinear optimization challenges.
Inverse-design approaches have been increasingly applied to magnonics, allowing researchers to automate the design of devices by specifying a desired functionality and using computational algorithms to find the optimal configuration\cite{papp2021nanoscale,wang2021inverse,yan2022inverse,zenbaa2024magnonic}.

In this paper, we present a novel discretization strategy for micromagnetic simulations, adjoint-state algorithms for efficiently solving time-dependent inverse problems, and the software design of NeuralMag, which integrates these advancements into a flexible and high-performance computational framework.

\section{Micromagnetics}
The micromagnetic model provides a semi-classical continuum description of magnetization dynamics in ferromagnetic systems, as originally formulated by Brown \cite{brown1959micromagnetics}.
The key governing equation is the Landau-Lifshitz-Gilbert (LLG) equation, which reads
\begin{equation}
  \frac{\partial \vec{m}}{\partial t}
  = -\frac{\gamma}{1 + \alpha^2} \vec{m} \times \vec{H}_\text{eff} - \frac{\alpha\gamma}{1 + \alpha^2}  \vec{m} \times (\vec{m} \times \vec{H}_\text{eff})
  \label{eq:llg}
\end{equation}
with $\vec{m}$ being the unit-vector field representation of the magnetization, $\gamma$ being the reduced gyromagnetic ratio, and $\alpha$ being a dimensionless damping parameter.
The effective field $\vec{H}_\text{eff}$ accounts for all relevant interactions within the system and derives from the total energy as
\begin{equation}
    \vec{H}_\text{eff}
    = -\frac{1}{\mu_0 M_\text{s}} \frac{\delta E}{\delta \vec{m}}
    \label{eq:heff}
\end{equation}
with $M_\text{s}$ being the saturation magnetization and $\delta E / \delta \vec{m}$ denoting the variational derivative of the energy with respect to the magnetization.
When the energy $E$ depends on spatial derivatives of the magnetization field $\vec{m}$, additional boundary conditions must be imposed to solve Eqs. \eqref{eq:llg} and \eqref{eq:heff}.
One such example is the micromagnetic exchange energy, which is defined as
\begin{equation}
    E_\text{ex} = \int_\Omega A (\vnabla \vec{m})^2 \dx,
    \label{eq:exchange_energy}
\end{equation}
where $A$ is the exchange stiffness constant.
The variation of the exchange energy with respect to $\vec{m}$ yields
\begin{equation}
    \delta E_\text{ex}(\vec{m}, \delta\vec{m}) =
    \int_\Omega \underbrace{- 2 [\vnabla \cdot (A \vnabla \vec{m})]}_{\equiv \delta E / \delta \vec{m}} \cdot \delta\vec{m} \dx + 
    \int_{\partial \Omega} \underbrace{2A  \frac{\partial \vec{m}}{\partial \vec{n}}}_{\equiv \vec{B}(\vec{n})} \cdot \delta\vec{m} \ds
    \label{eq:exchange_variation}
\end{equation}
leading to the exchange field definition
\begin{equation}
    \vec{H}_\text{ex}
    = -\frac{1}{\mu_0 M_\text{s}} \frac{\delta E_\text{ex}}{\delta \vec{m}}
    = \frac{2}{\mu_0 M_\text{s}} \vnabla \cdot (A \vnabla \vec{m}).
\end{equation}

The boundary term in Eq. \eqref{eq:exchange_variation} defines the appropriate exchange boundary condition.
To satisfy equilibrium conditions in micromagnetics, the system must fulfill Brown's conditions, which require $\vec{m} \times \delta E / \delta \vec{m} = 0$ for $\vec{x} \in \Omega$, and $\vec{m} \times \vec{B} = 0$ for $\vec{x} \in \partial \Omega$.

A similar variational treatment at internal interfaces, where material parameters vary discontinuously, introduces additional interface conditions \cite{abert2019micromagnetics}.
Assuming a continuous magnetization across such interfaces and dividing the domain into regions of continuous material parameters, the corresponding interface condition can be written as $\vec{m}_1 \times \vec{B}_1(\vec{n}) = \vec{m}_2 \times \vec{B}_2(\vec{n})$, where $\vec{B}_1(\vec{n})$ and $\vec{B}_2(\vec{n})$ represent the boundary terms on either side of the interface.

In case of the exchange energy being the only energy contribution introducing spatial derivatives and furthermore considering $\vec{m} \perp \partial \vec{m} / \partial{n}$, this leads to the well known exchange jump condition \cite{brown1959micromagnetics}
\begin{equation}
    A_1 \frac{\partial \vec{m}_1}{\partial \vec{n}}
    =
    A_2 \frac{\partial \vec{m}_2}{\partial \vec{n}}.
    \label{eq:exchange_jump}
\end{equation}
In addition to satisfying equilibrium conditions, the boundary and interface conditions must be consistently fulfilled at all times when solving the Landau-Lifshitz-Gilbert (LLG) equation \cite{abert2019micromagnetics}.

\section{Nodal Finite-Difference Scheme}\label{sec:nodal_fd}
Existing micromagnetic simulation software usually employs either a finite-difference discretization on regular grids \cite{miltat2007numerical,abert2019micromagnetics} or a finite-element discretization on irregular grids \cite{schrefl2007numerical,abert2019micromagnetics}.
The use of regular cuboid grids in the case of finite-difference micromagnetics allows for a very efficient computation of the demagnetization field by means of an FFT accelerated convolution.
On the other hand the finite-element method allows for the accurate modeling of complex structures due to the use of irregular meshes.

Moreover, finite-element micromagnetics provides a more subtle but sometimes highly relevant advantage over finite-difference micromagnetics:
In finite-element micromagnetics, the magnetization is usually explicitly defined on each mesh-vertex whereas standard finite-difference tools store one magnetization vector per simulation cell, which is typically taken to be the magnetization in the center of this cell.
While this difference appears to be insignificant for the micromagnetic modeling in the bulk, it plays a crucial role when considering material interfaces where the magnetization is subject to boundary and jump conditions.
\begin{figure}
	\includegraphics{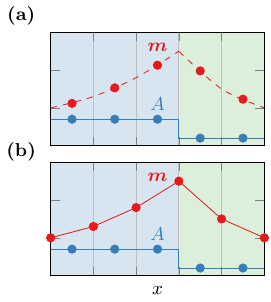}
    \caption{
        Illustration of the discretization of the magnetization $\vec{m}$ and the material parameter $A$ for a one-dimensional representation of a two-phase magnetic system, using different numerical schemes:
        (a) Standard finite differences: Both the material parameter and the magnetization are discretized with a single value per simulation cell. The magnetization degrees of freedom are treated as sample points of a continuous function.
        (b) Finite elements: The material parameters are discretized using piecewise constant functions, while the magnetization is represented as piecewise affine, with degrees of freedom located at the vertices.
    }
    \label{fig:fd_jump}
\end{figure}
Consider e.g. the exchange jump condition \eqref{eq:exchange_jump}, which prescribes a discontinuity in the first spatial derivative of the magnetization across material interfaces.
Choosing the degrees of freedom of the magnetization in the cell centers as illustrated in Fig.~\ref{fig:fd_jump}(a) requires a careful treatment of the boundary conditions that are defined on the vertices \cite{heistracher2022proposal}.
Similar considerations apply to interfacial energy contributions such as the RKKY coupling between two ferromagnetic layers \cite{suess2023accurate}.
Inaccurate modeling of the boundary conditions can lead to a loss of convergence order and consequently introduce significant numerical errors.
Introducing more energy contributions depending on surface integrals or spatial derivatives of $\vec{m}$ result in more complex boundary conditions \cite{abert2019micromagnetics} that become unfeasible to handle in standard finite-difference micromagnetics.

In contrast, the finite-element method allows for the choice of tailored function spaces for the magnetization and material parameters, as shown in Fig.~\ref{fig:fd_jump}(b).
Moreover, the inherently variational nature of the finite-element method allows to solve for the effective-field contributions by directly considering the variation of the energy \cite{abert2019micromagnetics} resulting in the weak form
\begin{equation}
    - \int_\Omega \mu_0 M_\text{s} \vec{H}(\vec{m}) \cdot \vec{v} \dx 
    =
    \delta E(\vec{m}, \vec{v})
    \quad \forall \quad \vec{v} \in V
    \label{eq:weak_form}
\end{equation}
with $V$ being a sufficiently smooth function space referred to as test space.
By a proper choice of function spaces for the material parameters and fields, this procedure does not require to explicitly account for the boundary conditions at all.

\subsection{Local Field Terms}
\begin{figure}
    \centering
    \includegraphics{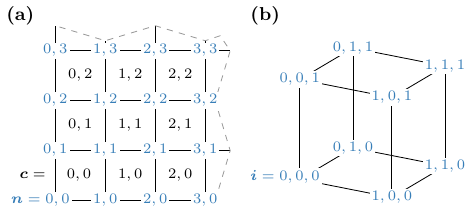}
    \caption{
        Cell and vertex numbering using multiindices for nodal finite-differences.
        (a) Two-dimensional representation of global cell and node indices denoted by $\vec{c}$ (black) and $\vec{n}$ (blue).
        (b) Three-dimensional representation of local vertex numbering denoted by index $\vec{i}$ (blue).
    }
    \label{fig:dof_numbering}
\end{figure}
The nodal finite-difference scheme proposed in this work applies the finite-element method for local field contributions on a regular cuboid grid.
This enables the use of an FFT accelerated demagnetization-field computation as in standard finite-difference micromagnetics, see Sec.~\ref{sec:demag}, while providing the rigoros and accurate handling of material interfaces for all local field contributions due to finite-element modeling.
In order to address the cells and nodes of the regular grid, we introduce multi-indices $\vec{c}$, $\vec{n}$ and $\vec{i}$ as
\begin{align}
    \vec{c} &= (c_1, c_2, c_3) \in \{0,\dots,N_1-1\}\times\{0,\dots,N_2-1\}\times\{0,\dots,N_3-1\}\\
    \vec{n} &= (n_1, n_2, n_3) \in \{0,\dots,N_1\}\times\{0,\dots,N_2\}\times\{0,\dots,N_3\}\\
    \vec{i} &= (i_1, i_2, i_3) \in \{0,1\}\times\{0,1\}\times\{0,1\}
\end{align}
with $N_1,N_2,N_3$ being the number of simulation cells in the respective mesh dimension.
The indices $\vec{c}$ and $\vec{n}$ are used to address simulation cells and nodes respectively according to the numbering introduced in Fig.~\ref{fig:dof_numbering}(a).
The index $\vec{i}$ either acts as local vertex number in a simulation cell according to Fig.~\ref{fig:dof_numbering}(b) or more general as a relative index to address neighborships.

We discretize all continuous fields appearing in weak forms with standard piecewise polynomial and globally continuous basis functions $\phi_{\vec{n}}$ that form a nodal basis on the cuboid mesh.
Each basis function $\phi_{\vec{n}}$ is defined per simulation cell in terms of reference basis functions $\hat{\phi}_{\vec{i}}$ as 
\begin{equation}
    \phi_{\vec{n}}(\vec{x}) =
    \sum_{\vec{i}}
    \hat{\phi}_{\vec{i}}\left[\begin{pmatrix}
        x_1 / \Delta x_1 - n_1 + i_1\\
        x_2 / \Delta x_2 - n_2 + i_2\\
        x_3 / \Delta x_3 - n_3 + i_3
    \end{pmatrix}\right]
    \label{eq:nodal_basis}
\end{equation}
with $\Delta x_k$ being the simulation-cell size in the $k$-th dimension.
The reference basis functions $\hat\phi_{\vec{i}}$ are defined on the reference unit cell $\Omega_\text{ref} = [0,1]\times[0,1]\times[0,1]$ as
\begin{align}
    \begin{split}
        \hat\phi_{\vec{i}}(\vec{x}) =
        \mathbbm{1}_{\Omega_\text{ref}}(\vec{x})
        &\big[ 1 - i_1 + (2 i_1 x_1 - x_1)\big] \cdot\\
        &\big[ 1 - i_2 + (2 i_2 x_2 - x_2)\big] \cdot\\
        &\big[ 1 - i_3 + (2 i_3 x_3 - x_3)\big]
    \end{split}
\end{align}
where $\mathbbm{1}_{\Omega_\text{ref}}$ denotes the characteristic function of $\Omega_\text{ref}$ which evaluates to 1 if $\vec{x} \in \Omega_\text{ref}$ and to 0 else.
This restricts the support of the reference basis functions $\hat\phi_{\vec{i}}$ to the reference cell $\Omega_\text{ref}$.
\begin{figure}
	\includegraphics{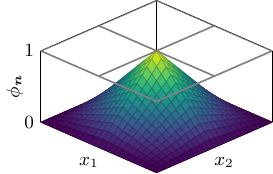}
	\caption{
        Two-dimensional representation of a basis function $\phi_{\vec{n}}$ in nodal finite-differences with a support spanning 4 simulation cells.
    }
    \label{fig:basis_function}
\end{figure}
A 2D representation of a basis function is visualized in Fig.~\ref{fig:basis_function}.
Furthermore, we introduce vector basis functions as
\begin{equation}
    \vec{\phi}_{\vec{n},j} = \phi_{\vec{n}} \vec{e}_j
\end{equation}
with $\vec{e}_j$ being the unit vector in direction $j \in \{1,2,3\}$.
Continuous vector fields such as the magnetization $\vec{m}$ and the effective field $\vec{H}_\text{eff}$ are then discretized as
\begin{equation}
    \vec{m} \rightarrow \vec{m}^\text{h} = \sum_{\vec{n},j} m_{\vec{n},j} \vec{\phi}_{\vec{n},j}
\end{equation}
with the superscript h denoting the discretized version of a field and coefficients $m_{\vec{n},j}$ being the nodal values of this field.

For material parameters, such as the saturation magnetization $M_\text{s}$, we choose a piecewise constant function space in order to allow for the accurate modeling of rapid material interfaces.
Namely, we define these parameters per simulation cell resulting in the following discretization
\begin{equation}
    M_\text{s} \rightarrow M_\text{s}^\text{h} = \sum_{\vec{c}} M_{\text{s},\vec{c}} \vartheta_{\vec{c}}
    \label{eq:ms_discrete}
\end{equation}
with basis functions
\begin{equation}
    \vartheta_{\vec{c}} = \mathbbm{1}_{\Omega_{\vec{c}}}.
    \label{eq:cell_basis}
\end{equation}
Replacing all fields with their discretized counterparts in the weak form \eqref{eq:weak_form} and testing with individual basis functions instead of arbitrary test functions yields the discretized weak form
\begin{equation}
    - \int_\Omega \mu_0 M_\text{s} \vec{H}^\text{h}(\vec{m}^\text{h}) \cdot \vec{\phi}_{\vec{n},j} \dx 
    =
    \delta E(\vec{m}^\text{h}, \vec{\phi}_{\vec{n},j})
    \quad \forall \quad \vec{n},j.
    \label{eq:weak_form_discrete}
\end{equation}

For a given node $\vec{n}$, we split the variation $\delta E(\vec{m}^\text{h}, \vec{\phi}_{\vec{n}, j})$ into its contributions from the 8 simulation cells that share node $\vec{n}$ and we address these cells by the local index $\vec{i} \in \{0,1\}^3$.
In general, the variation over a single simulation cell depends on the magnetization values of all nodes of this cell. 
Considering the three components of the magnetization, the contribution of the cell $\vec{i}$ to the variation can be written as
\begin{equation}
    \delta E^{\ast \vec{i}}_j = F_{\vec{i},j}(m_{\vec{i}',j'})
    \quad\text{for}\quad
    \vec{i}' \in \{0,1\}^3
    \text{ and }
    j' \in \{1,2,3\}
    \label{eq:weak_form_impl_f_cell}
\end{equation}
where $m_{\vec{i}',j'}$ denotes all nodal values of the magnetization in cell $\vec{i}$.
If the energy $E$ is quadratic in $\vec{m}$, the function $F$ is linear in $m_{\vec{i}',j}$ and can be described by a $24 \times 24$ matrix considering the $2^3 \cdot 3$ degrees of freedom defined by the index pairs $\vec{i},j$ and $\vec{i}', j'$.
In the finite-element context this matrix is usually referred to as element matrix of the weak form.

In order to compute the variation at all nodes, we introduce the vector $\vec{\delta E}$ with components $\delta E_{\vec{n}, j} = \delta E(\vec{m}^\text{h}, \vec{\phi}_{\vec{n}, j})$ and the auxiliary vectors $\vec{\delta E^{\ast \vec{i}}}$ containing the cell-wise variations according to \eqref{eq:weak_form_impl_f_cell} for all nodes.
Considering the node and cell numbering introduced in Fig.~\ref{fig:dof_numbering}, the global node index is given by the global cell index and the relative node index as $\vec{n}(\vec{c}, \vec{i}) = \vec{c} + \vec{i}$ resulting in
\begin{align}
    \delta E^{\ast \vec{i}}_{\vec{c} + \vec{i}, j} &=
    F_{\vec{i},j}(m_{\vec{c} + \vec{i}', j'})
    \quad\text{for}\quad
    \vec{i}' \in \{0,1\}^3
    \text{ and }
    j' \in \{1,2,3\}
    \label{eq:weak_form_impl_f}\\
    \vec{\delta{E}} &= \sum_{\vec{i}} \vec{\delta E^{\ast \vec{i}}} \label{eq:weak_form_impl}.
\end{align}
Note that $F$ only depends on the relative index $\vec{i}$ and the component $j$.
Eqs.~\eqref{eq:weak_form_impl_f} and \eqref{eq:weak_form_impl} deliver a straight-forward strategy for a parallel evaluation over the cell index $\vec{c}$, see Sec.~\ref{sec:form_compilation}.

If the energy $E$ depends on further fields, such as an external field or material parameters, the mapping function can be easily extended by adding additional arguments
\begin{equation}
    F_{\vec{i},j} (
        m_{\vec{c} + \vec{i}',j'},
        a^1_{\vec{c} + \vec{i}'},
        a^2_{\vec{c} + \vec{i}'},
        \dots, 
        b^1_{\vec{c}},
        b^2_{\vec{c}},
        \dots 
    )
\end{equation}
where the variables $a^l_{\vec{i} + \vec{j}}$ are the coefficients of arbitrary scalar fields discretized with nodal basis functions \eqref{eq:nodal_basis} and the variables $b^l_{\vec{i}}$ are the coefficients of arbitrary scalar fields discretized with cell basis functions \eqref{eq:cell_basis}.
Since $F_{\vec{j},k}$ does not explicitly depend on the cell index $\vec{i}$, it is fully determined by the integrand of the weak form \eqref{eq:weak_form_discrete} and the dimensions of a single simulation cell $\Omega_{\vec{i}}$.

In order to determine the discretized effective field $\vec{H}^\text{h}$, the weak form requires the solution of a linear mass system defined by the left-hand side of \eqref{eq:weak_form_discrete}.
To avoid this costly procedure, we employ mass lumping to the left-hand side of \eqref{eq:weak_form} as described in \citet{abert2019micromagnetics} resulting in

\begin{equation}
    H_{\vec{n},j}
    = - \left[ \int_{\Omega} \mu_0 M^\text{h}_\text{s} \phi_{\vec{n}} \dx \right]^{-1} \delta E_{\vec{n}, j}\\
    \label{eq:mass_lumping}
\end{equation}
where the saturation magnetization $M_\text{s}$ is discretized cell-wise according to \eqref{eq:ms_discrete}.

The proposed method is applicable to any energy contribution whose density depends solely on the magnetization and its first-order spatial derivatives, such as Zeeman energy, crystalline anisotropies, and both symmetric and antisymmetric exchange interactions.
Due to the regularity of the cuboidal grid, a matrix-free implementation of the presented scheme is straight-forward.
The local support of the basis functions results in a computational complexity of $\mathcal{O}(N)$ for the evaluation of any local field term with $N$ being the number of simulation cells.

\subsection{Demagnetization Field}\label{sec:demag}
To compute the demagnetization field, we employ the well-established FFT-accelerated method commonly used in standard finite-difference micromagnetic simulations \cite{miltat2007numerical}.
This algorithm calculates the demagnetization field generated by homogeneously magnetized cuboidal simulation cells arranged on a regular grid through fast convolution.
Since this method requires both the magnetization and the resulting field to be specified for each simulation cell, we introduce a straightforward pre- and post-processing step.
This procedure averages the values to transition between nodal and cell-centered discretizations efficiently.
FFT-accelerated methods that operate directly on node-wise discretized magnetizations have been proposed in previous studies\cite{berkov1993solving,ramstock1994optimizing}.
However, we opt for the standard method based on homogeneously magnetized cuboids due to its advantages in memory efficiency and computational performance specifically for 2D computations where the FFT also reduces to two dimensions.

\subsection{Low-Dimensional Geometries}\label{sec:low_dim}
\begin{figure}
    \centering
    \includegraphics{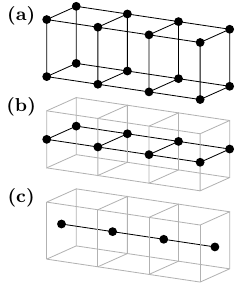}
    \caption{
        Representation of the degrees of freedom for a square-shaped rod using the following methods:
        (a) full three-dimensional discretization,
        (b) two-dimensional discretization with basis functions that are constant along the third dimension, and
        (c) one-dimensional discretization with basis functions that are constant along both the second and third dimensions.
    }
    \label{fig:low_dim}
\end{figure}
Discretizing a mesh with $N_1 \times N_2 \times N_3$ cells results in $(N_1 + 1) \times (N_2 + 1) \times (N_3 + 1)$ degrees of freedom when using a nodal basis for the function discretization.
In bulk system simulations, this introduces only a negligible overhead in comparison to standard finite-difference schemes, where the degrees of freedom are equal to the number of simulation cells. 
However, a significant application area for micromagnetic simulations involves magnetic thin films, which are often discretized with just a single layer of simulation cells.
In such cases, the 3D nodal discretization introduces a notable overhead -- roughly doubling the computational cost -- because it requires separate descriptions for the top and bottom surfaces of the thin film.
This contrasts with standard finite differences, where the problem effectively reduces to a 2D formulation.
By transitioning to 2D basis functions while maintaining full 3D integration in the weak form \eqref{eq:weak_form_discrete}, the nodal finite-difference scheme can accurately describe magnetic thin films.
This approach reduces the degrees of freedom to $N_1 \times N_2 \times 1$, making it more efficient for thin film simulations.
Namely, the 2D basis function on the reference cell are chosen as
\begin{align}
    \begin{split}
        \hat\phi_{\vec{i}}(\vec{x}) =
        \mathbbm{1}_{\Omega_\text{ref}}(\vec{x})
        &\big[ 1 - i_1 + (2 i_1 x_1 - x_1)\big] \cdot\\
        &\big[ 1 - i_2 + (2 i_2 x_2 - x_2)\big]
    \end{split}
\end{align}
with a 2D multiindex $\vec{i} = (i_1, i_2) \in \{0,1\}^2$.
As illustrated in Fig.~\ref{fig:low_dim} this approach can be also generalized to 1D problems leading to basis functions
\begin{align}
    \begin{split}
        \hat\phi_i(\vec{x}) =
        \mathbbm{1}_{\Omega_\text{ref}}(\vec{x})
        \big[ 1 - i + (2 i x_1 - x_1)\big]
    \end{split}
\end{align}
with a scalar index $i \in \{0,1\}$.

\section{Inverse Micromagnetics}
In addition to employing a nodal finite-difference scheme, NeuralMag is specifically designed to address inverse problems in both space and time domains.
In this context, the computation of individual field terms or the solution of the Landau-Lifshitz-Gilbert (LLG) equation \eqref{eq:llg} is classified as a forward problem $F$.
Given a vector of design variables $\vec{\theta}$, which may include material properties or the initial magnetization configuration, these forward problems yield well-defined outputs $\vec{y}$, such as effective field contributions or the resulting magnetization trajectory
\begin{equation}
    F(\vec{\theta}) = \vec{y}.
\end{equation}
An inverse problem is formulated to determine the design variables $\vec{\theta}$ that yield a specified result $\vec{y}$ from the forward problem.
This task is often challenging, as inverse problems are typically ill-posed, and their solution vectors may encompass a large number of degrees of freedom.
The most common strategy to solve such problem is the reformulation in terms of a minization problem
\begin{equation}
    \min_{\vec{\theta}} \mathcal{L}(\vec{\theta})
    \quad\text{with}\quad
    \mathcal{L}(\vec{\theta}) = \| F(\vec{\theta}) - \vec{y} \|
    \label{eq:inverse_min}
\end{equation}
that might be complemented by additional terms for regularization or smoothing purposes.
In the case of a high dimensional input $\vec{\theta}$ and a nonlinear function $F$ this problem is nontrivial.
In such cases, iterative methods, typically based on the gradient of the functional $\vnabla_{\vec{\theta}} \mathcal{L}$, are commonly employed to find a solution.
NeuralMag uses automatic differentiation \cite{paszke2017automatic} for static problems such as inverse strayfield calculations.
In contrast to the adjoint method that has been used in previous works \cite{bruckner2017solving,abert2017fast}, this approach performs the differentiation on the discrete level (discretize first).
As for the adjoint method, the gradient computation requires a forward solve and a subsequent backward solve with the complexity of the backward solve being equivalent to that of the forward solve.

For time-dependent problems, NeuralMag implements the adjoint-state method \cite{chen2018neural}.
The adjoint-state method is a powerful tool for the solution of PDE-constrained optimization problems also referred to as optimal-control problems.
Given a forward problem
\begin{equation}
    \frac{\partial \vec{m}}{\partial t} = \vec{f}_{\vec{\theta}}(t, \vec{m}),
    \label{eq:adjoint_forward}
\end{equation}
with design variables $\vec{\theta}$, we define an objective functional
\begin{equation}
    \mathcal L(\vec{\theta}) = L(\vec{m}(T;\vec{\theta}), \vec{y}_\text{target})
    \label{eq:adjoint_objective}
\end{equation}
with $\vec{m}(T;\vec{\theta})$ being the solution of \eqref{eq:adjoint_forward} for a final time $T$ and $\vec{y}_\text{target}$ being the desired output of the forward problem.
In order to compute the gradient of the objective functional with respect to the design variables $\vnabla_{\vec{\theta}} \mathcal L(\vec{\theta})$, the adjoint-state method requires two steps.
In the first step, the forward problem \eqref{eq:adjoint_forward} is solved for the given design variables $\vec{\theta}$, which results in the output $\vec{m}_\text{output} = \vec{m}(T)$.
In the second step, the so-called adjoint problem is solved, which is given by the following system of ODEs
\begin{equation}
    \begin{cases}
        \frac{\partial \vec{m}}{\partial t} = \vec{f}_{\vec{\theta}}(t, \vec{m})\\
        \frac{\partial \vec{a}}{\partial t} = - \frac{\partial \vec{f}_{\vec{\theta}}(t, \vec{m})}{\partial \vec{m}} \vec{a}\\
        \frac{\partial \vec{u}}{\partial t}       = - \frac{\partial \vec{f}_{\vec{\theta}}(t, \vec{m})}{\partial \vec{\theta}} \vec{a}
    \end{cases}
    \begin{array}{ll}
        \text{with} \quad \vec{m}(T) = \vec{m}_\text{output}\\
        \text{with} \quad \vec{a}(T) = \vnabla_y L(\vec{y}, \vec{y}_\text{target}), \vec{y} = \vec{m}_\text{output}\\
        \text{with} \quad \vec{u}(T) = \vec{0}.
    \end{array}
    \label{eq:adjoint_system}
\end{equation}
with $\vec{a}$ being the so-called adjoint variable.
This system is solved backwards in time starting from the final time $T$ used in the forward pass.
Successful integration of the system yields the output $\vec{u}(0)$ which can be identified as the desired gradient of the objective
\begin{equation}
    \vec{u}(0)=\vnabla_{\! \vec{\theta}} \mathcal L(\vec{\theta}).
\end{equation}
While the objective \eqref{eq:adjoint_objective} depends solely on the magnetization at the final time $T$, extending this method to objectives depending on multiple time points $T_i$ can be done in a straight-forward fashion by adding appropriate terms depending on $\vec{m}(T_i;\vec{\theta})$ to \eqref{eq:adjoint_objective}.
The computational and storage complexity of the adjoint system is comparable to that of the forward problem, yielding an exceptionally efficient strategy for the gradient computation of PDE-constrained optimization.
This method is superior to the backpropagation method \cite{wang2021inverse, papp2021nanoscale} with regard to the storage requirements that are similar to a regular forward pass.
However, this advantage comes at the cost of a reduced accuracy which is caused by the backwards pass that reconstructs the magnetization trajectory by inverse integration instead of using the exact values from the forward pass.

\section{Implementation}
NeuralMag is a Python library designed specifically for micromagnetic simulations, with a focus on high-performance tensor computations.
A key feature of NeuralMag is its ability to operate with either PyTorch\cite{paszke2019pytorch} or JAX\cite{frostig2018compiling,bradbury2021jax} as computational backend, allowing users to select the framework that best suits their needs.
By leveraging these modern machine learning frameworks, NeuralMag achieves efficient computations on a variety of parallel hardware, including CPUs, GPUs, and TPUs.
This versatility is complemented by the advantages these frameworks offer, such as optimized performance for large-scale simulations and built-in support for automatic differentiation, which simplifies solving inverse problems.
Through the modular design, the software is prepared to simplify the use of other computational backends in the future.

The use of either PyTorch or JAX as backends allows NeuralMag to fully exploit the unique strengths of each framework. PyTorch's \texttt{torch.compile()} feature enables just-in-time (JIT) compilation, optimizing the computational workflow by reducing operation overhead and enabling kernel fusion for faster execution on compatible hardware. However, PyTorch currently has limitations when compiling complex functions, such as those involving the demagnetization field, which means \texttt{torch.compile()} can only be applied to certain field terms.

In contrast, JAX’s \texttt{jit()} function can be applied to the entire right-hand side of the LLG equation.
This capability allows JAX to significantly reduce Python overhead and leads to notable performance gains, particularly for smaller systems where the overhead would otherwise be a bottleneck.

Both backends support single- and double-precision computations, enabling NeuralMag to offer users flexibility in balancing computational speed with numerical accuracy according to the requirements of each simulation.
The dual-backend approach ensures that NeuralMag can adapt to the user's preferred ecosystem while maintaining high computational efficiency and flexibility.

\subsection{Form Compilation}\label{sec:form_compilation}
\begin{lstlisting}[language=Python, caption={Symbolic definition the exchange energy \eqref{eq:exchange_energy} in NeuralMag.}, label={lst:energy}]
def e_expr(m, dim):
    A = Variable("material__A", "c" * dim)
    return (A * (
            m.diff(N.x).dot(m.diff(N.x)) +
            m.diff(N.y).dot(m.diff(N.y)) +
            m.diff(N.z).dot(m.diff(N.z))
        ) * dV(dim)
    )
\end{lstlisting}

\begin{lstlisting}[language=Python, caption={Automatically generated code for the computation of the exchange field.}, label={lst:form_code}]
def h(dx, m, material__A, material__Ms, rho):
    h = torch.zeros_like(m)
    h[:-1,:-1,:-1,0] += material__A[...]*rho[...]*(
        m[:-1,:-1,:-1,0]*(
            0.222222222222222*dx[0]*dx[1]/dx[2] +
            0.222222222222222*dx[0]*dx[2]/dx[1] +
            0.222222222222222*dx[1]*dx[2]/dx[0]) +
        m[:-1,:-1,1:,0]*(
            ...
        )
        ...
    )
    h[:-1,:-1,:-1,1] += ...
    h[:-1,:-1,:-1,2] += ...
    h[:-1,:-1,1:,0] += ...
	...
    return h / mass
\end{lstlisting}
At the heart of NeuralMag is a form compiler that translates a symbolic representation of a finite-element weak form into efficient tensor operations tailored to the chosen backend.
For symbolic computation, NeuralMag leverages the Python library SymPy \cite{meurer2017sympy}.
SymPy provides a powerful framework for representing the mathematical structures involved in micromagnetic simulations.
Specifically, NeuralMag introduces custom SymPy symbols to represent functions that are discretized either node-wise or cell-wise, as described in Sec.~\ref{sec:nodal_fd} of this paper.
Users can define the weak form of the micromagnetic problem using SymPy's symbolic language, allowing them to work in an intuitive mathematical formulation.

In addition to defining weak forms symbolically, NeuralMag leverages SymPy to automatically perform the variation of a symbolic energy expression, allowing it to derive the corresponding weak form.
This capability streamlines the process of converting complex energy functionals into their weak form representations.
For instance, in Lst.~\ref{lst:energy}, the exchange energy is defined symbolically using SymPy, demonstrating how users can express physical energy terms within the framework.

NeuralMag’s form compiler processes the symbolic weak form and transforms it into the discrete mapping function $F_{\vec{i},j}$, as defined in \eqref{eq:weak_form_impl_f}.
This transformation is achieved by applying Gauss quadrature to integrate over the finite elements, converting the weak form into a set of tensor operations—primarily multiplications and summations—that can be efficiently executed by the selected backend.
The role of the relative cell index, as discussed in equations \eqref{eq:weak_form_impl_f} and \eqref{eq:weak_form_impl}, is handled by tensor slicing.
This involves slicing along specific tensor dimensions by removing either the first \texttt{[1:]} or the last \texttt{[:-1]} value of the tensor in that dimension, which is necessary for handling the spatial relationships between adjacent cells in the discretized domain.
This systematic conversion of symbolic expressions into backend-specific tensor operations is key to NeuralMag's high-performance computational capabilities.
An example code snippet for the PyTorch backend, generated from the exchange energy defined in Lst.~\ref{lst:energy}, is shown in Lst.~\ref{lst:form_code}.
The generated function is highly optimized, as it operates solely on raw tensor objects without introducing any loops or conditional statements.
This structure ensures that the function is ideally suited for optimization by the just-in-time (JIT) compilers of both PyTorch and JAX.
By avoiding control flow statements, the generated code can be compiled into efficient low-level machine instructions, maximizing performance on parallel hardware architectures.

\subsection{Dynamic Attributes}
\begin{lstlisting}[language=Python, caption={Example usage of dynamic attributes in NeuralMag.}, label={lst:dynamic_attrs}]
>>> state = State(...)
>>> state.a = 1.0
>>> state.b = lambda a: 2.0 * a
>>> state.c = 5.0
>>> state.d = lambda b, c: b + c
>>> print(state.d)
7.0
\end{lstlisting}
\begin{lstlisting}[language=Python, caption={Automatically generated function for the evaluation of the dynamic attribute \texttt{d}.}, label={lst:dynamic_attrs_f}]
def lmda(a, c):
    b = __b(a)
    return __lmda(b, c)
\end{lstlisting}
NeuralMag introduces the concept of dynamic attributes through its state object, which allows attributes to be either tensors or functions that depend on tensors and return tensors.
This flexible design enables dynamic relationships between attributes, where some can be defined as functions of others, with NeuralMag automatically managing these dependencies.
For example, consider the code in Lst.~\ref{lst:dynamic_attrs}: attributes \texttt{a}, \texttt{b}, \texttt{c}, and \texttt{d} are defined, where \texttt{b} depends on \texttt{a}, and \texttt{d} depends on both \texttt{b} and \texttt{c}.
When \texttt{state.d} is accessed, NeuralMag resolves these dependencies, and the output is \texttt{7.0} because \texttt{d} is computed as the sum of \texttt{b} (which is \texttt{2 * a = 2.0}) and \texttt{c} (which is \texttt{5.0}).
Importantly, instead of scalar values, any tensor can be used as an attribute, allowing for more complex operations on multidimensional data. 

When defining such dynamic attributes, NeuralMag analyzes the function signatures to identify all dependencies in a recursive manner.
It then generates a new Python function at runtime that only relies on pure tensors and eliminates any control structures, such as loops or conditionals, ensuring the function remains optimal for high-performance tensor computation.
In the case of the example from Lst.~\ref{lst:dynamic_attrs}, the dynamically created function looks like Lst.~\ref{lst:dynamic_attrs_f}, where \texttt{d} depends on \texttt{b}.
Although \texttt{b} is not explicitly listed in the function arguments, its dependency on \texttt{a} is automatically resolved within the body of the function.
This approach simplifies the handling of complex dependencies, while maintaining the computational efficiency needed for the PyTorch and JAX backends.

\subsection{Automatic Differentiation and Time Integration}
In the context of inverse problems, NeuralMag leverages automatic differentiation and efficient time integration to solve complex optimization tasks.
Both PyTorch and JAX offer powerful automatic differentiation capabilities, which are crucial for computing gradients with respect to parameters in inverse problems.
For time integration, NeuralMag integrates with torchdiffeq\cite{kidger2021hey} (for PyTorch) and diffrax\cite{kidger2021on} (for JAX), both of which provide support for solving ordinary differential equations (ODEs).

Time integration is essential in dynamic micromagnetic problems, where the system's evolution must be accurately tracked.
Both libraries support a variety of numerical schemes for time stepping, including Euler methods, Runge-Kutta methods (such as RK4), and adaptive solvers like the Dormand-Prince method.
These methods ensure that NeuralMag can flexibly adapt to different accuracy and performance requirements in dynamic simulations.

For gradient-based optimization in inverse problems, NeuralMag supports both the adjoint method\cite{chen2018neural} and traditional backpropagation.
The adjoint method is particularly well-suited for problems with long time horizons or large state spaces, as it computes gradients more efficiently by solving an adjoint ODE backward in time.
Both torchdiffeq and diffrax support the adjoint method for time integration, offering an efficient way to compute gradients when optimizing over dynamic systems.
At the same time, they also allow for direct backpropagation through the time integration process, which can be more straightforward for shorter time intervals or simpler problems.

By combining automatic differentiation with advanced time integration techniques, NeuralMag can effectively tackle inverse problems in micromagnetic simulations, allowing users to optimize parameters while ensuring accurate numerical solutions over time.

\section{Validation and Benchmarks}
To validate the accuracy of NeuralMag, we solve two significant micromagnetic problems.
These tests showcase NeuralMag's ability to handle both standard and advanced cases, verifying its precision and computational efficiency.

\begin{figure}
    \centering
    \includegraphics{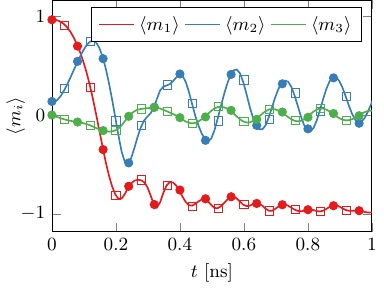}
    \caption{
        Results for MuMag Standard Problem \#4 are presented using both 2D and 3D discretizations as computed by NeuralMag. The reference solution, computed with OOMMF\cite{donahue1999oommf}, is depicted by solid lines for comparison. The NeuralMag solutions are illustrated using circles for the 3D discretization and squares for the 2D discretization.
    }
    \label{fig:sp4}
\end{figure}
The first validation case is MuMag Standard Problem \#4\cite{mumag4}, which simulates the dynamic behavior of a thin ferromagnetic film under an applied magnetic field.
The focus is on the time evolution of the averaged magnetization components.
We solve this problem using a full 3D spatial discretization and compare the results to a 2D simulation as described in Sec.~\ref{sec:low_dim} of the paper.
The results, displayed in Fig.~\ref{fig:sp4}, show excellent agreement with the reference solutions from the MuMag community, demonstrating the precision of NeuralMag in simulating the time dynamics of micromagnetic systems both with the 3D as well as 2D thin-film approximation.

\begin{table}
    \begin{tabular}{l|r|r|r}\toprule
        \multicolumn{1}{c|}{\textbf{discontinuous}}&
        \multicolumn{1}{c|}{\textbf{analytical}} &
        \multicolumn{1}{c|}{\textbf{magnum.af}} &
        \multicolumn{1}{c}{\textbf{NeuralMag}} \\
        \multicolumn{1}{c|}{\textbf{parameters}} &
        \multicolumn{1}{c|}{\textbf{[T]}} &
        \multicolumn{1}{c|}{\textbf{[T]}} &
        \multicolumn{1}{c}{\textbf{[T]}} \\ \midrule
        $A$/$K$/$M_\text{s}$ & 1.568 & 1.585 & 1.580\\
        $A$/$K$       & 1.089 & 1.116 & 1.112\\
        $A$/$M_\text{s}$     & 1.206 & 1.256 & 1.205\\
        $A$           & 0.838 & 0.868 & 0.867\\
        $K$/$M_\text{s}$     & 1.005 & 1.020 & 1.012\\
        $K$           & 0.565 & 0.582 & 0.571\\ \bottomrule
    \end{tabular}
    \caption{
        Depinning fields for a domain wall in a two-phase magnet, as defined in \cite{heistracher2022proposal}, computed with NeuralMag and compared to the analytical and numerical reference solutions computed with magnum.af.
    }
    \label{tab:sp_dw}
\end{table}
The second validation case involves solving the domain wall pinning problem proposed by Heistracher et al.\cite{heistracher2022proposal}.
This problem focuses on calculating the coercive field required to unpin a domain wall at the interface between two magnetic phases with varying material properties, such as exchange interaction, uniaxial anisotropy, and spontaneous magnetization.
This problem is sensitive to discontinuities in these parameters, making it an ideal test for NeuralMag’s handling of complex material boundaries.

In this validation, we compare the switching fields calculated by NeuralMag with the analytical results provided in Tab.~1 of the original paper.
We varied the material parameters (exchange constant $A$, anisotropy constant $K$, and saturation magnetization $M_\text{s}$) in different combinations across the two magnetic phases.
Tab.~\ref{tab:sp_dw} compares the switching fields obtained using NeuralMag with those presented in the reference paper.
Our results closely match the analytical solutions, with minor deviations likely due to the time integration method and field rate used during the simulation.
These successful validations confirm that NeuralMag correctly handles discontinuities at material interfaces and provides accurate predictions for complex micromagnetic systems.

\begin{figure}
    \centering
    \includegraphics{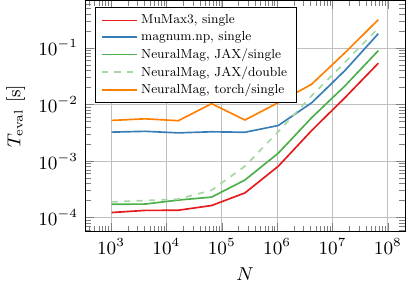}
    \caption{
        Comparison of the computation time for evaluating the right-hand side of the Landau-Lifshitz-Gilbert (LLG) equation, including both the exchange and demagnetization fields, across various system sizes $N$, in comparison with other finite-difference codes.
        The legend indicates the code as well as the floating-point precision used for the computation.
    }
    \label{fig:benchmark}
\end{figure}
To evaluate the performance of NeuralMag, we conducted a throughput benchmark, shown in Fig.~\ref{fig:benchmark}, where we compare the time required for evaluating the right-hand side (RHS) of the Landau–Lifshitz–Gilbert (LLG) equation across different system sizes.
Specifically, we measure the time for the integration of the full LLG including the exchange and demagnetization field and then divide by the number of field evaluations.
This procedure can be easily applied to any micromagnetic code without the need to modify it and provides a robust measure of the over-all performance at the same time.
In this benchmark, NeuralMag is compared to two widely-used micromagnetic simulation tools: mumax3\cite{vansteenkiste2014design} and magnum.np\cite{bruckner2023magnum}.
mumax3 shows the best performance due to its highly optimized GPU implementation.
However, NeuralMag, when using JAX as the backend, almost matches the performance of mumax3, being less than a factor of 2 slower.

Remarkably, NeuralMag maintains this competitive performance even for small system sizes, despite the computational overhead typically associated with a Python implementation.
This performance can be attributed to the just-in-time (JIT) compilation feature of JAX, which optimizes the entire RHS of the LLG equation at runtime.
Thanks to NeuralMag's architecture, JAX is able to analyze and compile the full computation into highly optimized machine code, reducing overhead and achieving near-optimal execution times.
This demonstrates the strength of NeuralMag’s design in leveraging modern machine learning frameworks to achieve high-performance computations while maintaining flexibility.

The remaining performance gap of approximately a factor of two compared to MuMax3 likely arises from the demagnetization field computation, as the current FFT interface in JAX is limited, preventing certain optimizations. However, as JAX’s FFT capabilities expand, this gap could narrow significantly --- or even vanish completely --- in the future.

\section{Inverse Problems}
\begin{figure}
    \centering
    \includegraphics{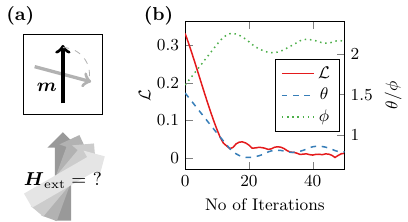}
    \caption{
        Simple inverse micromagnetic problem for the optimization of the external field direction in order to align the magnetization of a single-domain particle in a given direction.
        (a) Sketch of the problem setup.
        (b) Convergence of the objective function $\mathcal{L}$ and the optimized field angles $\theta$ and $\phi$.
    }
    \label{fig:inverse}
\end{figure}
\begin{lstlisting}[language=Python, caption={Simulation script for inverse problem.}, label={lst:inverse}]
state = State(...)
state.angles = [jnp.pi / 2, jnp.pi / 2]
h_ext = lambda angles: jnp.stack(
    [
        Hc / 2 * jnp.sin(angles[0]) * jnp.cos(angles[1]),
        Hc / 2 * jnp.sin(angles[0]) * jnp.sin(angles[1]),
        Hc / 2 * jnp.cos(angles[0]),
    ]
)
...
llg = nm.LLGSolver(state, parameters=["angles"])
m_target = nm.VectorFunction(state).fill((0.5**0.5, 0, 0.5**0.5)).tensor

def loss(angles, args):
    m_pred = llg.solve(state.tensor([0.0, 0.05e-9]), angles).ys[-1]
    return jnp.mean((m_target - m_pred) ** 2)

solver = optx.BFGS(1e-3, 1e-3, optx.max_norm)
result = optx.minimise(loss, solver, state.angles)
\end{lstlisting}
To demonstrate the solution of time-dependent inverse problems using NeuralMag, we solve a straightforward optimization problem.
Specifically, we aim to optimize the direction of an external magnetic field to align the magnetization of a single-domain particle with a target configuration $\vec{m}_\text{target}$ after a given time $T$, see Fig.~\ref{fig:inverse}(a).
The optimization minimizes the objective function $\mathcal{L}$ with respect to the field angles $\theta$ and $\phi$, as defined by the system
\begin{align}
    \vec{H}_\text{ext}(\theta, \phi) &= H_c \begin{pmatrix}
    \sin(\theta) \cos(\phi)\\
    \sin(\theta) \sin(\phi)\\
    \cos(\theta)
    \end{pmatrix},\\
    \vec{H}_\text{eff} &= \vec{H}_\text{aniso} + \vec{H}_\text{exchange} + \vec{H}_\text{ext},\\
    \mathcal{L}(\theta, \phi) &= \int_{\Omega} \|\vec{m}(T) - \vec{m}_\text{target}\| \dx
\end{align}
with $\vec{m}(t)$ being constrained by the LLG \eqref{eq:llg}.
A shortened code listing demonstrating the setup for this inverse problem is provided in Lst.~\ref{lst:inverse}.
In this simple optimization, convergence is achieved after 30-50 gradient-descent steps, see Fig.~\ref{fig:inverse}(b).
NeuralMag computes the gradient of the objective function by performing one forward and one backward simulation of the dynamic problem.

\section{Code Development and Availability}
The source code of NeuralMag is publicly available under the GNU LGPL License on GitLab at \url{https://gitlab.com/neuralmag/neuralmag}\cite{neuralmag2024}.
Comprehensive documentation, an API reference, and tutorials are available at \url{https://neuralmag.gitlab.io/}.
NeuralMag can be installed via standard package managers such as pip or conda.
Users and community members are encouraged to contribute to the codebase, tutorials, and documentation.
Continuous integration workflows are set up using GitLab CI/CD to automatically run tests after every code change.
These tests are run with both the PyTorch and the JAX backend and include unit, integration, and system tests, covering both the standard problems and benchmarks.
The repository includes all numerical problems discussed in this paper, as well as the code to reproduce the benchmarks.

\section{Conclusion}
In this paper, we have introduced NeuralMag, an open-source Python library for micromagnetic simulations that leverages modern machine learning frameworks such as PyTorch and JAX to achieve high performance.
NeuralMag implements a novel nodal finite-difference discretization scheme, which provides a rigoros numerical description of continuous fields such as the magnetization as well as discontinuous material parameters.
This approach is particularly useful for the accurate modeling of material interfaces, while maintaining the same computational complexity as standard finite-difference schemes.
Its performance is competitive with state-of-the-art micromagnetic simulation codes, yet it offers unparalleled flexibility due to its Python-based interface and support for optimized tensor operations on a variety of hardware platforms.

NeuralMag is especially well-suited for solving inverse problems, particularly those with time-dependent objectives, thanks to its ability to seamlessly compute gradients using automatic differentiation.
This makes it a powerful tool for a wide range of optimization and simulation tasks in micromagnetics.
NeuralMag is freely available\cite{neuralmag2024}, making it accessible to the broader research community for further development and application.

\section{Acknowledgements}
This research was funded in whole or in part by the Austrian Science Fund (FWF) 10.55776/P34671, 10.55776/I6068, and 10.55776/PAT3864023. For open access purposes, the author has applied a CC BY public copyright license to any author-accepted manuscript version arising from this submission.
We gratefully acknowledge the wedding of the Koraltans for fruitful discussions and great time which led to this publication.

\bibliography{refs}
\end{document}